\documentclass[aps,prc,preprintnumbers,twocolumn,superscriptaddress,showpacs,amsmath,amssymb]{revtex4}

\usepackage{amsmath,amssymb} % Enhanced mathematics
\usepackage{graphicx}% Include figure files
\usepackage{dcolumn}% Align table columns on decimal point

\usepackage{bm}% bold math

\begin{document}

\preprint{BNO-LNFI/09-11}

\title{Pulse Shape Analysis and Identification of Multipoint Events
in a Large-Volume Proportional Counter in an Experimental Search for
2K Capture $^{78}$Kr}

\author{Yu.M.~Gavriljuk}
% \altaffiliation[Also at ]{}%Lines break automatically or can be forced with \\
\author{A.M.~Gangapshev}%
\author{V.V.~Kazalov}%
\author{V.V.~Kuzminov}%
\affiliation{Baksan Neutrino Observatory INR RAS, Russia
}%
\author{S.I.~Panasenko}
\author{S.S.~Ratkevich}
\affiliation{V.N.Karazin Kharkiv National University, Ukraine
}%
\author{S. P. Yakimenko}%
\affiliation{Baksan Neutrino Observatory INR RAS, Russia
}%

\begin{abstract}
A pulse shape analysis algorithm and a method for suppressing the
noise component of signals from a large copper proportional counter
in the experiment aimed at searching for 2K capture of $^{78}$Kr are
described. These signals correspond to a compound event with
different numbers of charge clusters due to from primary ionization
is formed by these signals. A technique for separating single- and
multipoint events and determining the charge in individual clusters
is presented. Using the Daubechies wavelets in multiresolutional
signal analysis, it is possible to increase the sensitivity and the
resolution in extraction of multipoint events in the detector by a
factor of $3\div4$.
\end{abstract}


\maketitle

\section{\label{Intr}Introduction}
The main modern achievements in studying processes of double beta
decay ($\beta^-\beta^-$-decay) can be attributed to detection of its
two-neutrino mode. This process has been discovered in as many as
ten nuclei (see reviews
\cite{Zd2002_DNDT},\cite{Bar2002_Cz},\cite{Cremonesi06}). The data
obtained for the two-neutrino mode offer a chance to directly
compare different models of the nuclear structure, which form the
basis for calculations of nuclear matrix elements $\|M^{2\nu}\|$,
and to select the optimal one. Though direct correlation between the
values of nuclear matrix elements for the two-neutrino and
neutrinoless modes of $\beta\beta$ decay is absent lacking, the
methods for calculating $\|M^{2\nu}\|$ and $\|M^{0\nu}\|$ are very
close, and a chance possibility to estimate their accuracy in
calculating $\|M^{0\nu}\|$ appears only when comparing experimental
data and theoretical results calculations for the probability of
$2\beta (2\nu)$ decay.

It can be expected that acquisition of experimental data on the
other types of $2\beta$ transitions ($2\beta^+, \rm{K}\beta^+$, and
$2\rm{K}$ processes) will make it possible to considerably increase
the quality of calculations for both $2\nu$ and $0\nu$ decays. Much
efforts have been currently made in searching for these processes
(\cite{r3},\cite{r4},\cite{r5}) in spite of the fact that the
$2\beta^+(2\nu)$ and $K\beta^+(2\nu)$ modes are strongly suppressed
relative to $2\beta(2\nu)$ decay due to the Coulomb barrier for
positrons, and a substantially lower kinetic energy attainable in
such transitions. Positrons are absent in the final state of the
$2K(2\nu)$ decay, and the kinetic energy of the transition may be
rather high (up to 2.8 MeV), which dictates determines an increased
probability of a decay. However, this process is also difficult to
detect, since it is only characteristic radiation that is detectable
in it. The state-of-the-art experimental limit on the $^{78}$Kr
half-life period with respect to the $2K(2\nu)$ capture is $T_{1/2}
\le 1.5\times10^{21}$ yr (90\% C.L.) \cite{r6}. The theoretical
calculations based on different models predict the following
$^{78}$Kr half-lives for this process: $3.7\times10^{21}$ yr
\cite{r7}, $4.7\times10^{22}$ yr \cite{r8}, and $7.9\times10^{23}$
yr \cite{r9}. The last two values were obtained from the estimates
of the $^{78}$Kr half-life with respect to the total number of
$2e(2\nu)$ captures including in view of the 78.6\% fraction of
$2K(2\nu)$ capture events which makes 78.6\% \cite{r10}. From
comparison of the experimental and theoretical results, it is
apparent that the sensitivity of measurements has reached the lower
limit of theoretical estimates.

\section{\label{sec1}The technique of the experiment}
The $^{78}$Kr$(2e_\texttt{K},2\nu)^{78}$Se reaction produces a
$^{78}$Se$^{**}$ atom with two vacancies   in its $K$-shell. The
technique for seeking this reaction is based on the assumption that
the values of   energies of characteristic photons, and of the
probability that they will be emitted when the double vacancy is
being filled, coincide with the   corresponding values   when two
separate single vacancies   in the $K$ shells of two isolated singly
ionized Se$^*$ atoms are being filled. In this case, the total
measured energy is $2K_{ab} = 25.3$ keV, where $K_{ab}$ is the
binding energy of a $K$ electron in a Se atom (12.65 keV). The
fluorescence yield  upon filling of a single vacancy   in the
$K$-shell of Se is 0.596. The energies and relative intensities of
the characteristic lines in the $K$ series are $K=11.22$ keV
(100\%), $K=11.18$ keV (52\%), $K=12.49$ keV (21\%), and $K=12.65$
keV (1\%) \cite{r11}. There are three  possible ways  for
deexcitation of a doubly ionized $K$-shell: 1) emission of  Auger
electrons only ($e_a$, $e_a$), 2) emission of a single
characteristic quantum and an Auger electron ($K,e_a$), and  3)
emission of two characteristic quanta and low-energy Auger electrons
($K,K,e_a$) , with probabilities $p1=0.163$, $p2=0.482$, and
$p3=0.355$, respectively. A characteristic quantum   can travel a
long enough distance in a gas medium between the points of its
production and absorption. For example, 10\% of characteristic
quanta with energies of 11.2 a nd 12.5 keV is absorbed in krypton at
a pressure of 4.35 atm ($\rho=0.0164$ g/cm$^3$) on a path 1.83 and
2.42 mm long, respectively (the values of absorption factors   are
taken from \cite{r12}). The  paths of photoelectrons   with the same
energies are 0.37 and 0.44 mm, respectively. They produce almost
pointwise charge clusters of primary ionization in the gas.  In case
of the event with the escape of two characteristic quanta   absorbed
in the working gas and a single Auger electron, the energy will be
distributed among three pointwise charge clusters. It is these
three-point (or three-cluster) events possessing a unique set of
features that were the subject of the search in \cite{r6}.

\begin{figure}[!htb]
\resizebox{0.45\textwidth}{!}{%
  \includegraphics{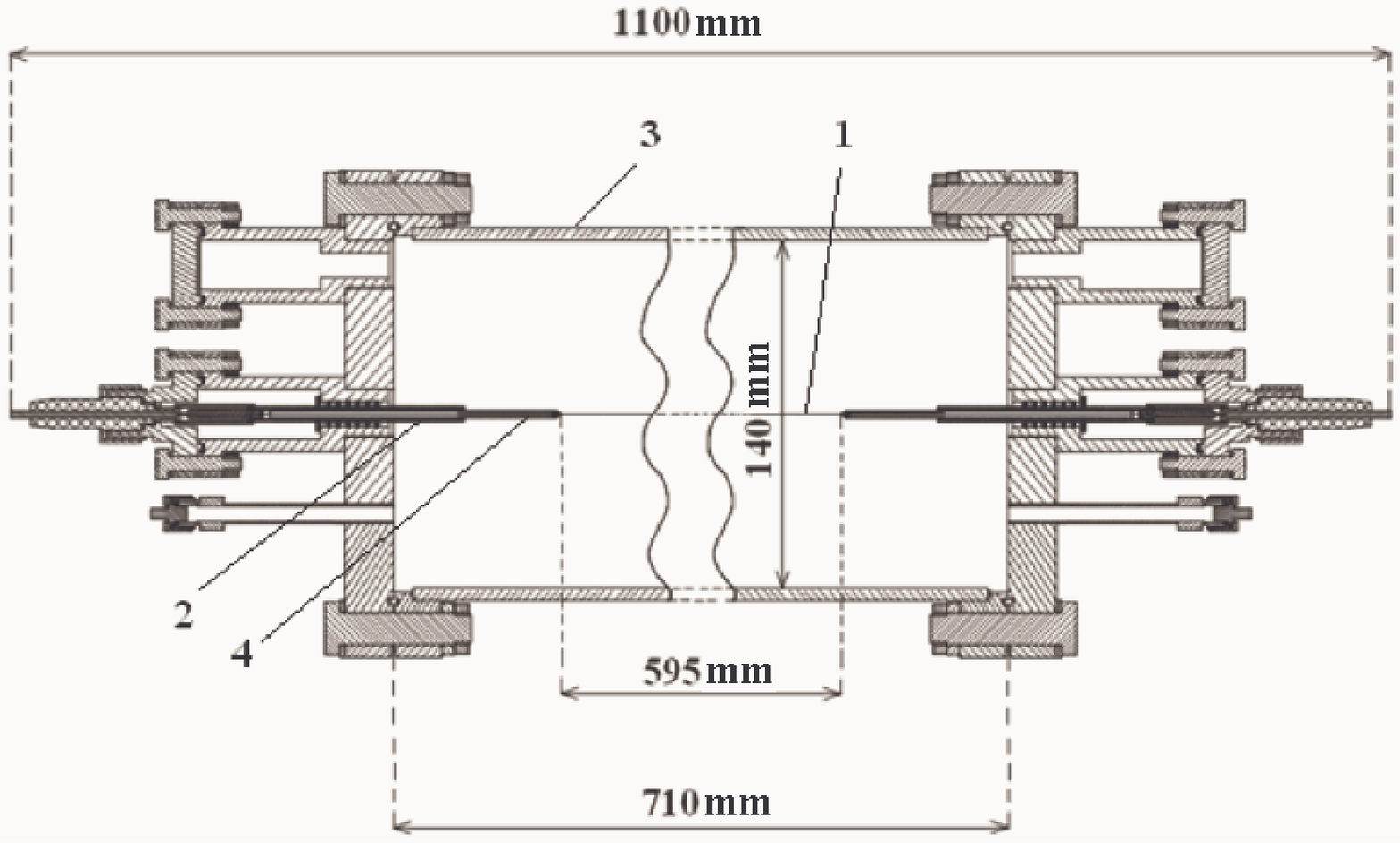}}
\caption{ Schematic view of the LPC in section along the anode wire:
1 -- wire (collecting electrode), 2 -- load-carrying insulator, 3 --
cathode, 4 -- tubular bulge of the anode.} \label{LPC}
\end{figure}

A large proportional counter (LPC) with a casing made of M1-grade
copper is used to detect   the above considered processes. The LPC
has a cylindrical shape with inner and outer diameters of 140 and
150 mm, respectively; its section along the axis is schematically
shown in Fig.\ref{LPC}. A gold-plated tungsten wire of 10 $\mu$m in
diameter goes along the LPC axis   and  serves as the anode. The
potential of +2400 is applied to the wire,   and the casing (the
cathode) is grounded. Both ends of the anode are lead to the
appropriate end cap flanges via high-voltage pressure-sealed
bushings-ceramic insulators with a central electrode taken from
spark plugs.

To reduce the influence of edge effects on the operating
characteristics of the counter, the end segments of the wire are
passed through the copper tubes with dimensions of $3.0\times38.5$
mm, which are electrically connected to the anode. Gas amplification
is absent on these segments, and charges are collected in an
ionization mode. With   the fluoroplastic insulator, the distance
from the working region to the flange is 70 mm.

The working part of the LPC is 595 mm in length (the distance
between the end caps of the tubes); therefore, the LPC s operating
volume is 9,159 l. The total capacitance of the counter and the
outlet insulator is $~30.6$ pF. The total resistance of the anode
and two output electrodes is $~600$ Om. All detachable joints are
sealed with indium wire. All nipple joints are sealed with
fluoroplastic gaskets. The internal insulators are made of
fluoroplastic. Their thickness was selected so as to be the smallest
possible in order to improve the degassing conditions during vacuum
treatment of the counter and stabilize its operating characteristics
in the course of measurements.

The LPC is filled with a pure Kr sample to a total pressure of 4.51
att; no quenching or accelerating gases are added. Prior to filling,
Kr is purified of electronegative impurities in a Ni/SiO$_2$
reactor.

The LPS's signals are read out by a charge-sensitive amplifier (CSA)
from one end of the anode wire. The CSA parameters have been
selected so that the signal is transmitted with minimum distortions,
and information of the spatial distribution of primary-ionization
charges in a projection onto a counter radius is fully represented
by the pulse shape. When amplified in an auxiliary amplifier, the
pulses arrive at the input of the   digital oscilloscope
LA-n20-12PCI, the output data of which (the pulse waveform digitized
with a frequency of 6.25 MHz) are recorded   with a personal
computer. The length of the scanning frame with a resolution of 160
ns is 1024 points (163.8 $\mu$s), of which $~50$ $\mu$s is the
"prehistory"  and $~114$ $\mu$s is the "history".

\begin{figure}[!htb]
\resizebox{0.45\textwidth}{!}{%
  \includegraphics{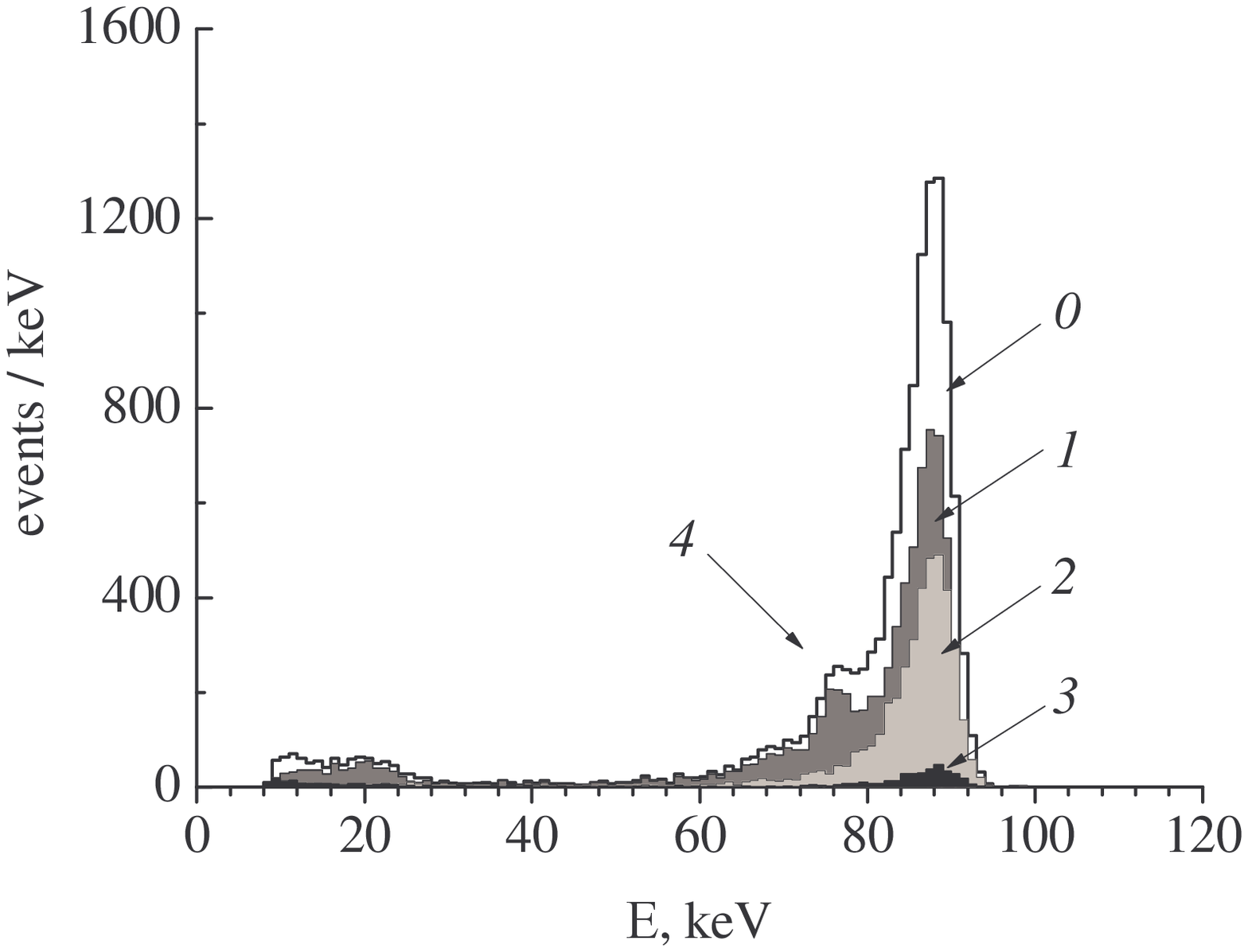}}
\caption{ Pulse amplitude spectrum from the external $^{109}$Cd
source located in the middle of the LPC length: (\emph{0}) all
events, (\emph{1}) single-point events, (\emph{2}) two-point events,
(\emph{3}) three-point events, and (\emph{4}) escape peak of
characteristic $\gamma$-ray photons.} \label{pic2}
\end{figure}

The counter is calibrated through the wall of its casing   by
$\gamma$ rays of a $^{109}$Cd source ($E_\gamma=88$ keV; relative
yield 0.036 photons/decay). Figure \ref{pic2} presents ({\it0}) the
total pulse amplitude spectrum and the energy spectra of ({\it1})
single-, ({\it2}) two-, and ({\it3}) three-point events from the
source located in the middle of the LPC length. The procedure for
obtaining them from digitized pulses is described in what follows.
The following factors make their contribution to the low-energy part
of the spectrum: characteristic radiation Ag$_{K_{\alpha\beta}}$
($E\approx22$ keV) from this source, which "survived"  after passing
through a 5-mm-thick copper wall; scattered radiation from the wall,
which is in equilibrium with the characteristic radiation, and
Compton electrons from scattering of 88-keV photons in the gas with
the escape of a Compton photon beyond the counter.

The 88-keV peak is wider on the low-energy side due to the
contribution of 88-keV $\gamma$ rays scattered from the wall. The
energy resolution of this peak, determined by its right half, is
6.5\%. Peak \emph{4} at an energy of 75.4 keV corresponds to the
escape of Kr characteristic radiation ($E_{Kr_\alpha}=12.6$ keV)
beyond the counter.

The 88-keV full-energy peak contains events with different internal
structures. Quanta with this energy are   absorbed in Kr mostly by
photoeffect   in the $K$-shell (86.7\%). The photoeffect in other
shells makes 13.3\% \cite{r12}, \cite{r13}. Filling of the vacancy
in the $K$-shell of Kr is accompanied by emission of characteristic
radiation in 66.0\% of cases and Auger electrons in 34.0\%
\cite{r13}. The theoretical efficiency of characteristic radiation
absorption in the counter s working volume is 86.9\%. Therefore, the
photoeffect is responsible for 49.7\% of two-point events
$(0.867\times0.660\times0.869)$ and 42.8\% of single-point events
$(0.133+0.867\times0.340)$ out of the total number of absorptions
due to photoeffect in the full-energy peak. By single-point events,
we mean all events in which only electrons escaping from the shell
of a single atom (photoelectrons $+$ Auger electrons), including
events of photoelectric absorption in the upper shells of Kr. Only
single-point events in an amount of 7.5\%
$[0.867\times0.660\times(1.000-0.869)]$ of the total number of
photoabsorption events will be presented in peak {\it4}.

Some primary quanta can be absorbed as a result of two-step process
of  "Compton scattering-photoeffect".  A Compton electron creates
one ionization point. A Compton photon absorbed by photoeffect
participates in the above-described processes. Therefore, the
two-step process makes its contributions to the full-energy peak in
the form of two- and three-point events and to peak {\it4} in the
form of two-point events. Upon normalization to the peak area, the
estimated final composition of events for the full-energy peak
contains 44.1\% (single-point events) + 51.2\% (two-point events)
due to photoeffect + 2.2\% (two-point events) + 2.5\% (three-point
events) due to the two-step process. In peak {\it4}, there are
95.3\% single-point events + 4.7\% two-point events.

\begin{figure}[!htb]
\resizebox{0.5\textwidth}{!}{%
  \includegraphics{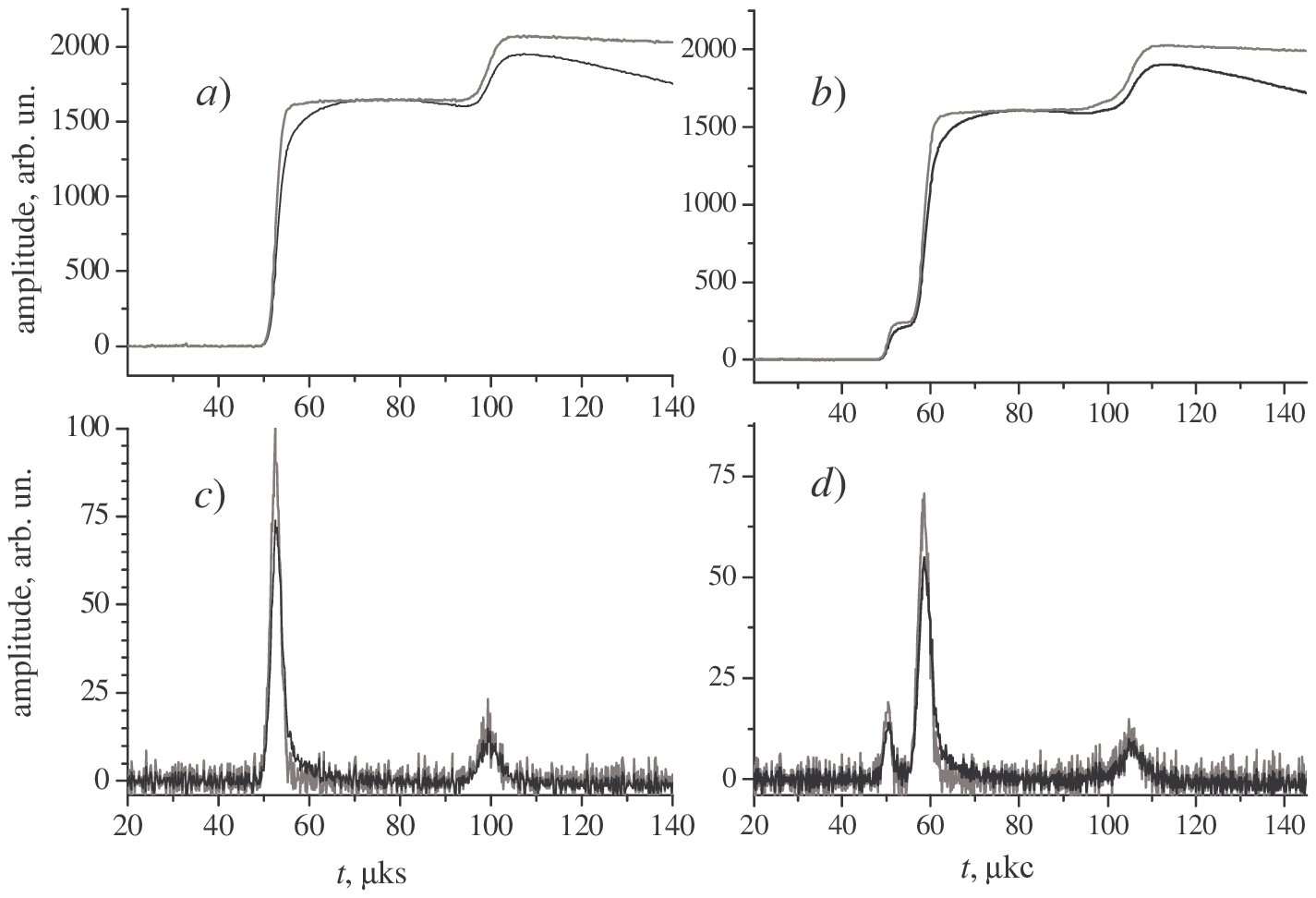}}
\caption{ Examples of pulses (dark lines) of two types of events:
(\emph{a}) for photoabsorption of an 88-keV photon with an escape of
electrons only (a single-point event), (\emph{b}) simultaneous
escape of a 12.6-keV characteristic photon Kr$_{K\alpha}$ and a
photoelectron (${\rm E}_\gamma -{\rm E}_{KrK\alpha}$)=88.0 keV--12.6
keV = 75.4 keV (a two-point event). The calculated, area-normalized
current pulses of primary ionization electrons are shown with light
lines in Figs. 3\emph{c} and 3\emph{d}, while the respective voltage
(charge) pulses obtained by integrating these current pulses are
depicted with light lines in Figs.3\emph{a} and 3\emph{b}.}
\label{exam_pulse}
\end{figure}
Examples of pulses (dark  lines) corresponding to events of two
types are presented in Fig.\ref{exam_pulse}. The pulse due to
photoabsorption of a 88-keV photon with escape of electrons only (a
single-point event) is shown in Fig.\ref{exam_pulse}a, and
simultaneous escape of characteristic photon Kr with an energy of
12.6 keV and a photoelectron $(E_\gamma-E)=88$~keV - 12.6 keV=75.4
keV (a two-point event) is illustrated in Fig.\ref{exam_pulse}b. The
maximum distance between pointwise charge clusters in projection
onto the counter s radius is equal to the radius. For pure Kr, the
evaluated time it takes for ionization electrons to drift from the
cathode to the anode is 53 $\mu$s. From Figs.\ref{exam_pulse}a and
\ref{exam_pulse}b, it is apparent that the second pulse with a
$\sim3\div5$ times smaller amplitude is produced in the counter in
about $\sim53$ $\mu$s after the first one. This pulse is generated
by secondary photoelectrons knocked out of the cathode by photons
produced during development of an avalanche from the primary
ionization. The probability of photoeffect   on the cathode is
rather high, since the working gas contains no quenchers.

\section{\label{shape} Determining the shape of the LPC current
signal}
The digitized output pulse from the measuring channel can be
presented as direct convolution of sought signal $x(t_i)$ with
response pulse $\widehat{G}$ from the linear system, which is
distorted by stochastic or determinate noise $z(t_i)$:
\begin{eqnarray}\label{Eq1}
  y(t_i )&:=& \widehat{G}x(t_i ) + z(t_i ), i = 0,...,N - 1\\
  &:=& (g \otimes x)(t_i ) + z(t_i ).
  \nonumber
\end{eqnarray}

The noise level determines the lower limit on the sensitivity of the
measuring channel, while the instrumental function defines the
resolution value.

Given the values of $y$ and $\widehat{G}$, one can try to estimate
$x(t)$ in the presence of noise $z(t)$; i.e., in principle, it is
possible to state the inverse problem (deconvolution) of determining
the signal at the linear system output by the values of the output
signal:
\begin{eqnarray}\label{Eq2}
\widetilde{x}(t_i ) = \widehat{G}^{ - 1} y(t_i ) = x(t_i ) +
\widehat{G}^{ - 1} z(t_i ),
\end{eqnarray}
where $\widehat{G}^{ - 1}$ is the operator inverse.

To determine the values of charges released in individual clusters
of a multipoint event, one can differentiate the original charge
pulse $\left( {\widehat{G}^{ - 1}  = \frac{d}{{dt}}} \right)$ and
represent the obtained shape by a set of Gaussian curves. The
evaluated area under an individual Gaussian curve will correspond to
the charge (energy) value in the relevant cluster. From
Figs.\ref{exam_pulse}\emph{c} and \ref{exam_pulse}\emph{d}, it is
apparent that direct differentiation provides  an asymmetric bell
shape (dark line). Such shape results from the nearly Gaussian
distribution of the current pulses due to electrons of primary
ionization from a pointwise  energy deposit, which arrive at the
boundary of the gas amplification region near the anode wire.

This shape is determined by the spatial distribution of the charge
density in projection onto the radius. The parameters of this
distribution depend on the time it takes for the primary charge
cluster to drift to the anode. As it drifts, the charge cluster
spreads out into a cloud due to electron diffusion. The pulse read
out from the anode wire is mostly produced by a negative charge
induced on the anode moving toward the cathode by positive ions
produced near the wire in gas amplification process and moving
toward the cathode i.e., the ion component (i.c.). The estimated
total ion drift time is 0.447 s. The contribution of the equilibrium
(with ions) electron component (e.c.) to the total induced charge is
$\sim7$\%. The electron collection time is $\sim1$ ns.

The output pulse shape is defined by the superposition of induced
charges from single electron avalanches distributed in time and
intensity according to: the shape of the current pulse from primary
ionization electrons, the shape of a pulse from an individual
avalanche, and a finite time of the CSA self-discharge. The last two
parameters are responsible for the asymmetry of the output current
pulse. The output current pulse can be transformed to a symmetric
shape by taking into account the analytical dependence of the
amplitude of the output voltage pulse generated by a point (in
projection onto the radius at the boundary of the gas amplification
region) group of primary ionization electrons as a function of time
and discharge constant of the output storage capacitor \cite{r14}:
\begin{eqnarray}
 V_{\emph{k}} (t_i ) &=& K_{{\rm{i.}}{\rm{c.}}} n_\emph{k} \exp \left( {
- \frac{{t_i  + B}}{{RC}}} \right) \times\\
& & \times\left\{ {ln\left( {1 + \frac{{t_i }}{\emph{B}}} \right) +
\frac{{t_i }}{{RC}} +
\frac{{t_i ^2 }}{{2 \cdot 2!(RC)^2 }} + ...} \right\}+ \nonumber\\
 & & +
K_{{\rm{e.}}{\rm{c.}}} n_\emph{k} \exp \left( { - \frac{{t_i
}}{{RC}}} \right),\nonumber
\end{eqnarray}
where $V_k(t_i)$ is the amplitude of the voltage pulse from the
$k$th group of electrons, $n_\emph{k}$ is the number of primary
electrons in the $k$th group, $t_i=t(t_0 + t_{i0})$ is the current
time for the dependence of the voltage pulse amplitude from the
$k$th group, $t_0$ is the time of origin of the total pulse, and
$t_{i0}$ is the time of origin of the pulse from the $k$th group,
$K_{\rm{i.c.}}= MV_{(1)i.c.}$, $M$ is the gas amplification factor,
\begin{eqnarray}
V_{(1){\rm i.c.}}=(e/C)\cdot\ln({\emph{r}}_{\rm{k}}/{\emph{r}}_0)
\ln({\emph{r}}_{\rm{k}}/{\emph{r}}_{\rm{a}})\nonumber
\end{eqnarray}
is the total voltage pulse amplitude produced at output capacitor
$\emph{C }$  by a single ion generated in the gas discharge,
$\emph{e}$ is the electron charge, $\emph{r}_0$ is the radius
corresponding to the avalanche's center of gravity, $\emph{r}_a$ is
the anode radius, $\emph{r}_{\rm{k}}$ is the cathode radius,
\emph{B} is the time parameter associated with the motion of
positive ions of the gas discharge in a particular gas (for the LPC
filled with Kr at 4.51 at, $\emph{B} = 2.28$ ns), $RC=\tau_d$ is the
CSA discharge constant, $R$ is the leakage resistance, $K_{\rm
{e.c.}}= MV_{(1){\rm{e.c.}}}$, and
\begin{eqnarray}
V_{(1){\rm
e.c.}}=(\emph{e}/\emph{C})\cdot\ln(\emph{r}_0/\emph{r}_{\rm{a}})
\ln(\emph{r}_{\rm{k}}/\emph{r}_{\rm{a}})\nonumber
\end{eqnarray}
is the total voltage pulse amplitude produced at output capacitor
$\emph{C }$ by a single electron generated in the gas discharge. In
Eq.(3), the electron component is assumed to appear instantly.

At $\tau_d=\infty $, Eq. (3) assumes the form
\begin{equation}\label{Eq4}
V_{\emph{k}} (t_i ) = K_{{\rm{i.}}{\rm{c.}}} n_\emph{k} \times
ln\left( {1 + \frac{{t_i }}{B}} \right) +
K_{{\rm{}}{\rm{e.}}{\rm{c.}}} n_\emph{k}.
\end{equation}
In our case, $\tau_d\approx192$ $\mu$s. For a time interval
satisfying condition $t/\tau_d < 2$, Eq. (3) can be reduced to the
first two expansion terms.

If its is assumed that the gas discharge from $n_k$ of primary
electrons happens at the beginning of the digitization interval, the
pulse amplitude at the end of this interval can be described by Eq.
(3) for $t=160$ ns, since the influence of the output capacitor
discharge over this time is negligible. If a contribution of the
earlier discharges is absent in this time interval, the pulse
amplitude at the upper bound of the interval can be used to
determine the $n_k$ value.  In this case, it is taken into account
that, at the end of a 160-ns interval, the contributions of the
terms to the total $V_k(t_i)$ value in Eq. (3) make 74 and 26\%,
respectively. These conditions are satisfied in the recorded actual
pulse in the first time channel from the beginning of the pulse. The
$n_1$ value obtained from the actual pulse is used in Eq. (3) to
calculate the total shape of the partial pulse in the entire time
interval from the beginning to the end of the frame. The pulse
obtained thereby is subtracted from the actual one. Therefore, the
above condition is now fulfilled for the first digitization interval
of the residual pulse or for the second interval of the original
pulse. This procedure is repeated until the last time channel in the
frame.

The sequence of $n_k$ values for a single-point event has a
symmetrical distribution with a nearly Gaussian shape. It is this
distribution that is used for further analysis. The area under the
Gaussian curve or, in the case of a multiparticle event, the sum of
the Gaussian areas on a time interval of 53 $\mu$s from the
beginning of the pulse yields the total number of primary ionization
electrons. To plot the spectra in Fig.2, this sum is multiplied by
the coefficient equal to the averaged ratio of areas of the actual
current pulse and of the calculated Gaussian curve for purely
single-point events.

The calculated, area-normalized current pulses of primary ionization
electrons at the boundary of the gas amplification region are shown
with light curves in Figs.\ref{exam_pulse}\emph{c} and
\ref{exam_pulse}\emph{d}, and the corresponding voltage (charge)
pulses obtained by integrating these current pulses are depicted
with light lines in Figs \ref{exam_pulse}\emph{a} and
\ref{exam_pulse}\emph{b}.

From Figs.\ref{exam_pulse}\emph{c} and \ref{exam_pulse}\emph{d}, it
is apparent that, at an energy deposit of 88 keV, the
signal-to-noise ratio (SNR) is rather high. In the energy range of
$20\div30$ keV, in which the $2K$-capture in $^{78}$Kr is sought,
the SNR for individual components of the total energy deposit
corresponding to the possible effect (25.3 keV) is not so favorable.

\begin{figure}[htp]
\resizebox{0.45\textwidth}{!}{%
  \includegraphics{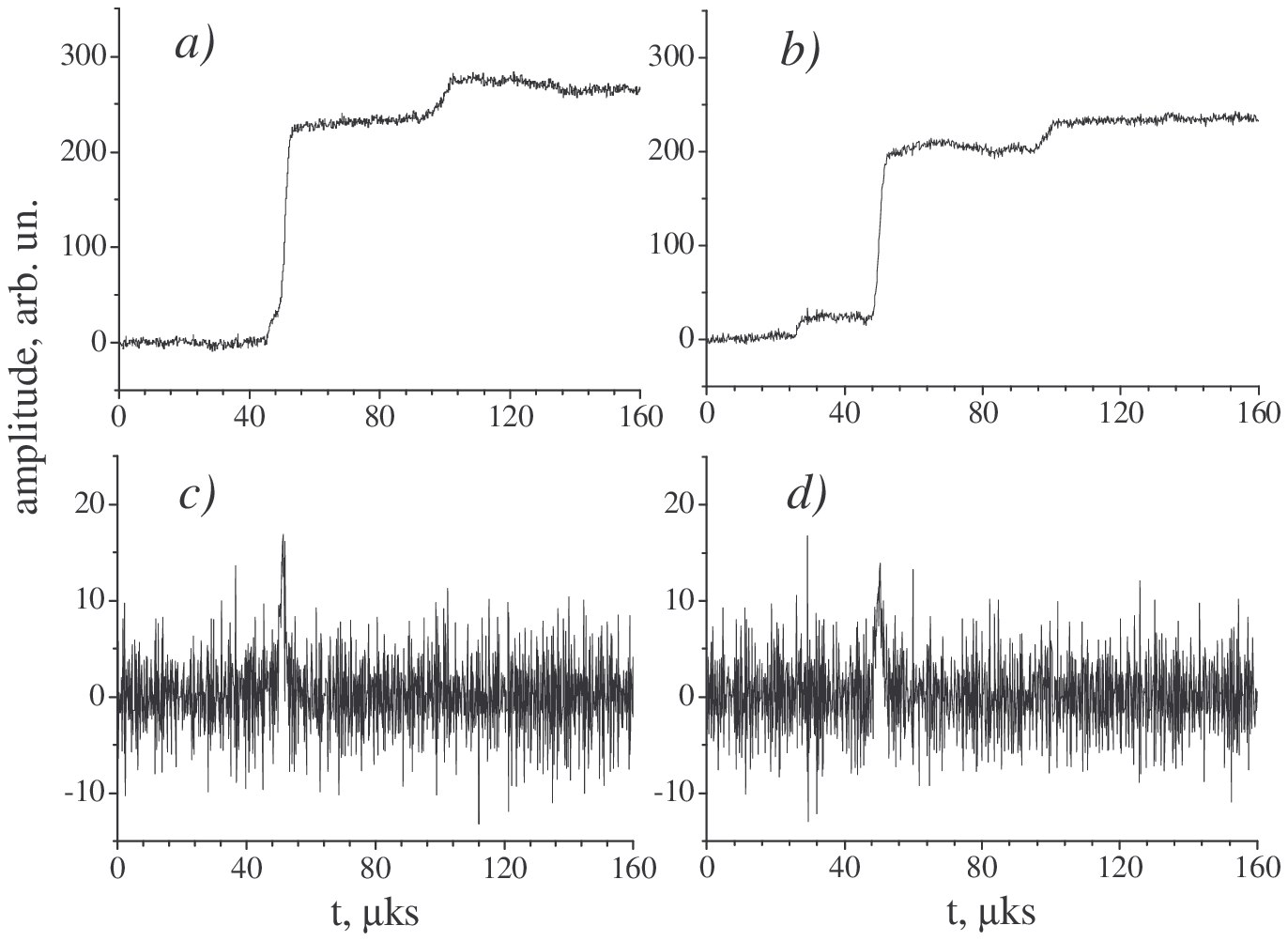}}
\caption{Primary ionization signals for two types of two-point
energy deposits from $K$-capture of $^{81}$Kr isotope accompanied by
an escape of a characteristic photon and a concomitant Auger
electron: (\emph{a}, \emph{b}) recalculated voltage (charge) pulses
and (\emph{c}, \emph{d}) corresponding current pulses from primary
ionization electrons.} \label{Pic4}
\end{figure}

Figures \ref{Pic4}\emph{a} and \ref{Pic4}\emph{b} present examples
of recalculated current pulses for two types of two-point energy
deposits due to the $K$-capture of $^{81}$Kr isotope. The total
energy deposit corresponds to the binding energy of an electron in
the $K$-shell of a daughter $^{81}$Br (13.5 keV). The energies of a
characteristic quantum ($E_{K\alpha}=11.9$ keV) and concomitant
Auger electrons ($E_a=1.6$ keV) are close to the energies of
individual components for events of the $2K$-capture of $^{78}$Kr.
More detailed information on the $^{81}$Kr source is given in what
follows.

From Fig.\ref{Pic4}, it is apparent that these signals have a high
noise level. Noises and possible electric pickup may both mask the
low-energy component and create a false one. The use of traditional
methods of frequency filtering with different window functions,
e.g., in the form of the Hamming \cite{r15}, Wiener \cite{r16}, and
Savitzky-Golay \cite{r17} filters, sometimes fail to ensure reliable
extraction of closely spaced (Fig. 4a) and masking each other
components of a compound event.

\section{\label{filters} Wavelet-based filter}

Mathematical studies carried out in late 1980s initiated intense
development of a principally new class of orthogonal transforms
based on the use of wavelet functions \cite{r18}. Wavelet transforms
are distinguished by a high degree of locality of their base
functions both in the time and frequency regions, which allows one
to use them for processing of many nonstationary processes. "In the
preliminary processing of our LPC data it is expedient to use
multiresolutional signal analysis \cite{r19} based on the {\it
dyadic transform of discrete signals}, and often called discrete
wavelet transform (DWT). In this case, the analyzed signal
$\widetilde{x}(t)$ is presented as the decomposition
\begin{equation}\label{Eq5}
x^J (t) = \sum\limits_{m = 0}^{N_{j_o }  - 1} {\widehat{a}_{j_0 ,m}}
\varphi_{j_0,m}(t) + \sum\limits_{j = j_0 }^J {\sum\limits_{m =
0}^{N_j  - 1}{\widehat{d}_{j,m} \psi_{j,m}}} (t),
\end{equation}
where are the well-known orthonormal scaling (scaling functions) and
wavelet functions- "ancestor" wavelets; $\widehat{a}_{j_0 ,m}  =
\left\langle {\widetilde{x},\varphi _{j_o ,m} } \right\rangle $ are
the empirical approximation coefficients and $\widehat{d}_{j,m}  =
\left\langle {\widetilde{x},\psi _{j,m} } \right\rangle $ are the
empirical detailing coefficients; $j,m\in Z$ are the current values
of the scale and the shift; $N_{j_0}(N_j)$ is the number of
approximation (detailing) coefficients considered at the relevant
levels of the decomposition; $j_0$ is the initial scale value; and
$J$ is the final scale value.

Parameter $J$ specifies the resolution of wavelet reconstruction
$x^J(t)$ of original signal $x(t)$. Actually, in the case of
tendency $J \rightarrow \infty$, the norm tends to $\|x^J-x\|
\rightarrow 0$.

Scaling functions $\{\phi_{j,m}(t)\}$ and mother wavelet functions
$\{\phi_{j,m}(t)\}_{j,m\in{Z}}$, which have $2K$ nonzero
coefficients, satisfy the so-called two-level relations \cite{r18}:
\begin{equation*}
\varphi (t) = \sqrt 2 \sum\limits_{n = 0}^{2K - 1} {h(n)\varphi (2t
- n)} ,
\end{equation*}
\begin{equation*}
\psi (t)=\sqrt 2 \sum\limits_{n = 0}^{2K-1}{g(n)\varphi (2t-n)} ,
\end{equation*}
where $h(n)$ and $g(n)$ are the coefficients of low- and
high-frequency filters of the wavelet transform, when $g(n) = \left(
{ - 1} \right)^n h(2K - n - 1)$. As distinct from other types of
transforms, in which the base functions are explicitly specified,
one succeeds in analytically obtaining the base functions in the
wavelet analysis only in rare cases, and the basis is most
frequently specified by coefficients $h(n)$ and $g(n)$. In this
paper, we use wavelets of the Daubechies family. The scaling
functions and the Daubechies wavelets are continuous functions that
are not identically equal to zero on a finite segment and are not
differentiable anywhere on this segment. Table \ref{tab1} contains
the filter coefficients used in the Daubechies scaling functions
$db4$ and $db6$ (numbers 4 and 6 denote the number of nonzero
coefficients in the filters). They are rational numbers and fully
define the Daubechies wavelet transform (DVT) \cite{r18}.
\begin{table} \label{tab1}
\caption{Coefficients of the low-frequency filters for wavelets with
compact carriers \emph{db}4 and \emph{db}6. The \emph{h}(\emph{n})
values are normalized so that $\sum\nolimits_{n=0}^{2K -
1}{h(n)=\sqrt 2}$ \cite{r18} }\label{tab1}
\begin{tabular}{|c|c|c|c|c|c|c|}
  \hline
  \emph{h}(\emph{n)} & $db4$ $(K=2)$ &  $db6$ $(K=3)$ \\
  \hline \hline
  \emph{h}(0) & ${{\left( {1 + \sqrt 3 } \right)} \mathord{\left/
 {\vphantom {{\left( {1 + \sqrt 3 } \right)} {4\sqrt 2 }}} \right.
 \kern-\nulldelimiterspace} {4\sqrt 2 }}$
 & ${{\left( {1 + \sqrt {10}  + \sqrt {5 + 2\sqrt {10} } } \right)} \mathord{\left/
 {\vphantom {{\left( {1 + \sqrt {10}  + \sqrt {5 + 2\sqrt {10} } } \right)} {16\sqrt 2 }}} \right.
 \kern-\nulldelimiterspace} {16\sqrt 2 }}$ \\
  \emph{h}(1) & ${{\left( {3 + \sqrt 3 } \right)} \mathord{\left/
 {\vphantom {{\left( {3 + \sqrt 3 } \right)} {4\sqrt 2 }}} \right.
 \kern-\nulldelimiterspace} {4\sqrt 2 }}$
 & ${{\left( {5 + \sqrt {10}  + 3\sqrt {5 + 2\sqrt {10} } } \right)} \mathord{\left/
 {\vphantom {{\left( {5 + \sqrt {10}  + 3\sqrt {5 + 2\sqrt {10} } } \right)} {16\sqrt 2 }}} \right.
 \kern-\nulldelimiterspace} {16\sqrt 2 }}$
 \\
  \emph{h}(2) & ${{\left( {3 - \sqrt 3 } \right)} \mathord{\left/
 {\vphantom {{\left( {3 - \sqrt 3 } \right)} {4\sqrt 2 }}} \right.
 \kern-\nulldelimiterspace} {4\sqrt 2 }}$
& ${{\left( {10 - 2\sqrt {10}  + 2\sqrt {5 + 2\sqrt {10} } }
\right)} \mathord{\left/
 {\vphantom {{\left( {10 - 2\sqrt {10}  + 2\sqrt {5 + 2\sqrt {10} } } \right)} {16\sqrt 2 }}} \right.
 \kern-\nulldelimiterspace} {16\sqrt 2 }}$
\\
  \emph{h}(3) & ${{\left( {1 - \sqrt 3 } \right)} \mathord{\left/
 {\vphantom {{\left( {1 - \sqrt 3 } \right)} {4\sqrt 2 }}} \right.
 \kern-\nulldelimiterspace} {4\sqrt 2 }}$
 & ${{\left( {10 - 2\sqrt {10}  - 2\sqrt {5 + 2\sqrt {10} } } \right)} \mathord{\left/
 {\vphantom {{\left( {10 - 2\sqrt {10}  - 2\sqrt {5 + 2\sqrt {10} } } \right)} {16\sqrt 2 }}} \right.
 \kern-\nulldelimiterspace} {16\sqrt 2 }}$
 \\
 \emph{h}(4) & -- & ${{\left( {5 + \sqrt {10}  - 3\sqrt {5 + 2\sqrt {10} } } \right)} \mathord{\left/
 {\vphantom {{\left( {5 + \sqrt {10}  - 3\sqrt {5 + 2\sqrt {10} } } \right)} {16\sqrt 2 }}} \right.
 \kern-\nulldelimiterspace} {16\sqrt 2 }}$
 \\
  \emph{h}(5) & -- & ${{\left( {1 + \sqrt {10}  - \sqrt {5 + 2\sqrt {10} } } \right)} \mathord{\left/
 {\vphantom {{\left( {1 + \sqrt {10}  - \sqrt {5 + 2\sqrt {10} } } \right)} {16\sqrt 2 }}} \right.
 \kern-\nulldelimiterspace} {16\sqrt 2 }}$
 \\
  \hline
\end{tabular}
\end{table}
In contrast both to the Fourier transform, which localizes
frequencies, but fails to provide time resolution for the process,
and to the apparatus of $\delta$-functions, which localizes moments
of time, but have no frequency resolution, the DVT makes it possible
to reveal the local properties of any structure of an individual
event on different scales, eliminating smooth polynomial
characteristics and emphasizing fluctuation structures. It is
possible to completely eliminate statistical fluctuations by
selecting only strong correlated fluctuations, which will allow one
to observe exactly those dynamic fluctuations that exceed the
statistical component (noise).

\subsection{Noise elimination using the wavelet
method}

Among the modern noise suppression techniques, the Donoho-Johnstone
method has received the widest acceptance \cite{r20}. This method is
rather simple in implementation and time-saving in computational
aspect, since it implies the use of only fast algorithms of wavelet
transform. It consists of three steps, which, being successively
applied to the original signal, produce a noise-suppression effect.
At the first step, signal under investigation $x(t)$ is subjected to
DWT; afterward, the threshold noise elimination procedure is applied
to each of the detailing coefficients of level $j$ and, sometimes,
to the approximation coefficients of the same level; and, finally,
inverse wavelet transform is performed, which results in
reconstruction of the signal that is characterized, as expected, by
a higher SNR.

This technique is a nonparametric estimate of the regression signal
model with the use of an orthogonal basis \cite{r21}-\cite{r23}; it
operates with high efficiency with signals in the decomposition of
which only a few detailing coefficients significantly differs from
zero. Selection of a particular wavelet form depends on the problem
under investigation and is not predetermined beforehand \cite{r24}.
The depth of decomposition generally depends on the properties of
the analyzed signal. Smooth wavelets produce a smoother signal
approximation, and vice versa- "short"  wavelets better search for
peaks of the approximated function. The depth of decomposition
affects the scale of rejected details; i.e., as the depth of
decomposition increases, the model subtracts the noise of a steadily
increasing level until the scale of details becomes too large  and
the transform starts distorting the original signal shape. Upon
further increase in the depth of decomposition, the transform starts
forming a smoothed version of the original signal; i.e., apart from
the noise, some local peculiarities of the original signal are also
filtered off. Decomposition of the signal and its reconstruction by
the approximation and detaining coefficients that have passed
threshold processing is carried out using the Malla algorithm
\cite{r19} and the lifting procedure \cite{r25}. The threshold
processing itself is performed with the aid of one of the threshold
noise suppression operations. In accordance with the {\it hard
threshold} processing \cite{r20}, all coefficients $\left\{
{\widehat{d}_{j,k} \left| {k \in Z} \right.} \right\}$ of level $j$,
which are greater than or equal to the threshold, are held constant,
while the other coefficients that do not satisfy this condition are
are reduced to zero:
\begin{equation}\label{Eq6}
    d_{j,k}^h  = \widehat{d}_{j,k} I\left( {\left| {\widehat{d}_{j,k} } \right| > \theta _j } \right),\begin{array}{*{20}c}
   {} & {\theta _j  = \rho _j \sigma _i ,}  \\
\end{array}
\end{equation}
where $\rho_j$ is the threshold factor for the specified scale
\cite{r19}, and $\sigma$ is the noise variance  on the $j$th scale.
Hard threshold processing is a straight-out procedure. It should
nevertheless be taken into account that a hard threshold has two
drawbacks that can lower its usefulness for the noise suppression
task. The first consists in the fact that retaining of detailing
coefficients above a predetermined threshold value also implies
retaining of their noise. The other drawback is in the presence of
parasitic harmonics generated in the resulting signal by artificial
introduction of lacunas (gaps) formed due to the coefficients
reduced to zero. The use of {\it soft threshold} processing implies
recalculation of detaining coefficients $d$ in the following way:
\begin{equation}\label{Eq7}
d_{j,{}^{}k}^s  = {\rm{sign}}\left( {\widehat{d}_{j,{}^{}k} }
\right)\left( {\left| {\widehat{d}_{j,{}^{}k} } \right| - \theta
_{trh} } \right),
\end{equation}
where $\theta$ is a certain threshold value.

In this case, apart from the reduction to zero of the coefficients
$d$ actually containing only the noise component, the detailing
coefficients are decreased by the $\theta_{thr}$ value, which
corresponds to noise suppression in the informative coefficients as
well. Threshold $\theta_{thr}$, which defines the analyzed signal,
is a sole parameter that cannot be exactly estimated directly from
actual data. Mathematical solutions in determining the threshold
value seem unsatisfactory to be a priori  accepted. In an ideal
case, we tend to specify this threshold so as to minimize the
root-mean-square error (rmse):
\begin{equation}\label{Eq8}
\varepsilon  = \sqrt {\frac{1}{N}\sum\limits_{i = 1}^N {\left[
{x(t_i ) - \widetilde{x}(t_i )} \right]} ^2 } .
\end{equation}

Therefore, by analogy to \cite{r19},\cite{r25}, we determine the
basic signal component relying on threshold processing of the model
signal. Ideal threshold processing minimizes the rmse, making all
coefficients with the signal component below $\sigma$ vanish
\cite{r19}. In other words, it is assumed that the free from noise
signal component above $\sigma$ is an original signal; i.e.,
$\theta_{thr}=\sigma$. This selection can be tested on the model
signals. Our experience gained in working with natural signals
confirms that the ratio of the signal peaks to the noise level
reaches its maximum near $\theta_{thr} = \sigma$.

\subsection{Test of model signal with noise}
 To illustrate the efficiency of different approaches to the
deconvolution problem, let us select model signal $x(t_n)$, which is
specified on a set of points $N=2^{10}$ and consists of three
closely spaced Gaussian peaks
\begin{equation*}
  a_k \exp \left( { - {{(t - t_k )^2 } \mathord{\left/
 {\vphantom {{(t - t_k )^2 } {2\mu _0^2 }}} \right.
 \kern-\nulldelimiterspace} {2\mu _0^2 }}} \right),
\end{equation*}
where $a_k=[1.32, 7.0, 7.0]$, $t_k=[277, 307, 327]$, and $\mu_o=7$.
Let us add Gaussian noise and a harmonic term to it:
\begin{equation*}
\widetilde{x}(t_n ) = x(t_n ) + \kappa \Delta (\tau ) + {\rm{g}}(t_n
),
\end{equation*}
where $\Delta (\tau)$ is ordinary Brownian process analyzed on
$\tau\in[0,1]$, $\kappa$ is a certain coefficient, and $g(t_n)$ is
the harmonic term.

For the variance $\sigma^2$ of the noise component in model signal
$W(t_n ) = \kappa \Delta (\tau ) + {\rm{g}}(t_n )$
 to be estimated by the $x(t_n)$ data, it is necessary
that the influence of signal $x(t_n)$ be suppressed. A rough
estimate can be obtained from the mean values of the smallest-scale
wavelet coefficients \cite{r20}. The signal with length $N$ has
$N/2$ wavelet coefficients $\left\{ {\left\langle {\widetilde{x}(t_n
),\psi _{j,m} } \right\rangle } \right\}_{0 \le m \le {N
\mathord{\left/
 {\vphantom {N 2}} \right.
\kern-\nulldelimiterspace} 2}}$ of the smallest scale $2^j=2N^{-1}$.
Coefficients $\left| {\left\langle {x(t_n ),\psi _{j,m} }
\right\rangle } \right|$ are small if signal $x(t_n)$ is smooth on
carrier $\psi_{j,m}$, and in this case, $\left\langle
{\widetilde{x}(t_n ),\psi _{j,m} } \right\rangle  \approx
\left\langle {W,\psi _{j,m} } \right\rangle $. However, if $x(t_n)$
has a sharp differential on carrier $\psi$, coefficients $\left|
{\left\langle {x(t_n ),\psi _{j,m} } \right\rangle } \right|$
 are rather large. In the case of a piecewise-smooth signal, we have several
differentials, which results in a certain number of large
coefficients. This number is small in comparison with $N/2$. At the
smallest scale, signal $x(t_n)$ defines the value of a small portion
of large-amplitude coefficients $\left| {\left\langle {x(t_n ),\psi
_{j,m} } \right\rangle } \right|$, which are considered to be
"overshoots".  All the other coefficients are approximately equal to
$\left\langle {W,\psi _{j,m} } \right\rangle $; they are independent
random Gaussian variables with variance $\sigma^2$. Therefore,
$\sigma^2$ can be roughly estimated, with the influence of $x(t_n)$
being ignored, by the mean of absolute values $\left\{ {\left\langle
{\widetilde{x}(t_n ),\psi _{j,m} } \right\rangle } \right\}_{0 \le m
\le {N \mathord{\left/
 {\vphantom {N 2}} \right.
 \kern-\nulldelimiterspace} 2}} $, divided  by 0.6745 \cite{r21}.

Daubechies filters $db4$ were used in a direct wavelet transform and
when the estimate of the useful signal was synthesized from
coefficients after their threshold processing, whereas higher-order
filters $db6$ were employed to estimate the threshold values.

Calculations of all wavelet coefficients, testing of the noise
suppression algorithms, and selection of required signals for
further analysis were performed using MATLAB and its web libraries
\cite{r26}.

\begin{figure}[!htb]
\resizebox{0.4\textwidth}{!}{%
  \includegraphics{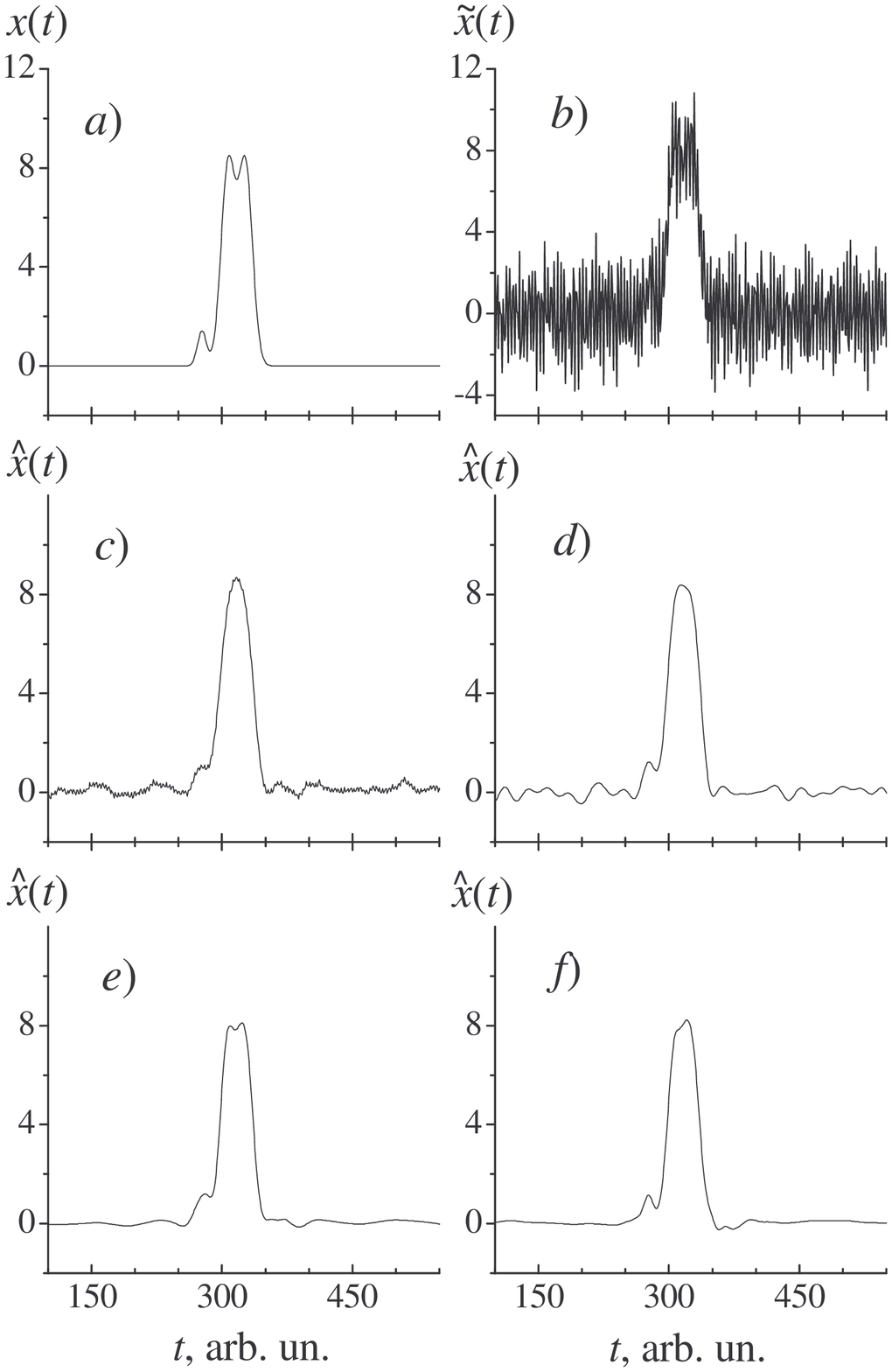}}
\caption{ (\emph{a}) Model signal from three closely spaced Gaussian
peaks; (\emph{b}) additive mixture of the model signal with noise;
noise suppression using (\emph{c}) the Savitzky--Golay filter and
(\emph{d}) Wiener filter; (\emph{e}, \emph{f}) Daubechies wavelet
filters, $db4$, with hard and soft threshold processing,
respectively.} \label{test_pulse}
\end{figure}
The results are shown in Fig.\ref{test_pulse}. It is apparent that
all filters provide good noise suppression efficiency (the SNR is
$>30$ dB), but the signals passed through the wavelet filters with
the use of the hard or soft threshold processing better represents
the shape of the original signal (Figs.\ref{test_pulse}\emph{e} and
\ref{test_pulse}\emph{f}), which ensures more reliable signal
resolution into individual components.  For our purposes, hard
threshold processing is quite satisfactory both in the quality
factor and in the computational time.

\section{\label{reliability} Reliability in identifying multipoint
events}

In order to verify the efficiency of the algorithm for the noise
suppression and pulse shape discrimination for multipoint events
from an actual source the LPC was filled with natural Kr containing
cosmogeneous radioactive isotope $^{81}$Kr ($T_{1/2}=2.1\times10^5$
yr) with a bulk activity of $\sim0.1$ $min^{-1}l^{-1}$Kr \cite{r27},
\cite{r28}.
\begin{figure}[!h]
\resizebox{0.4 \textwidth}{!}{%
\includegraphics{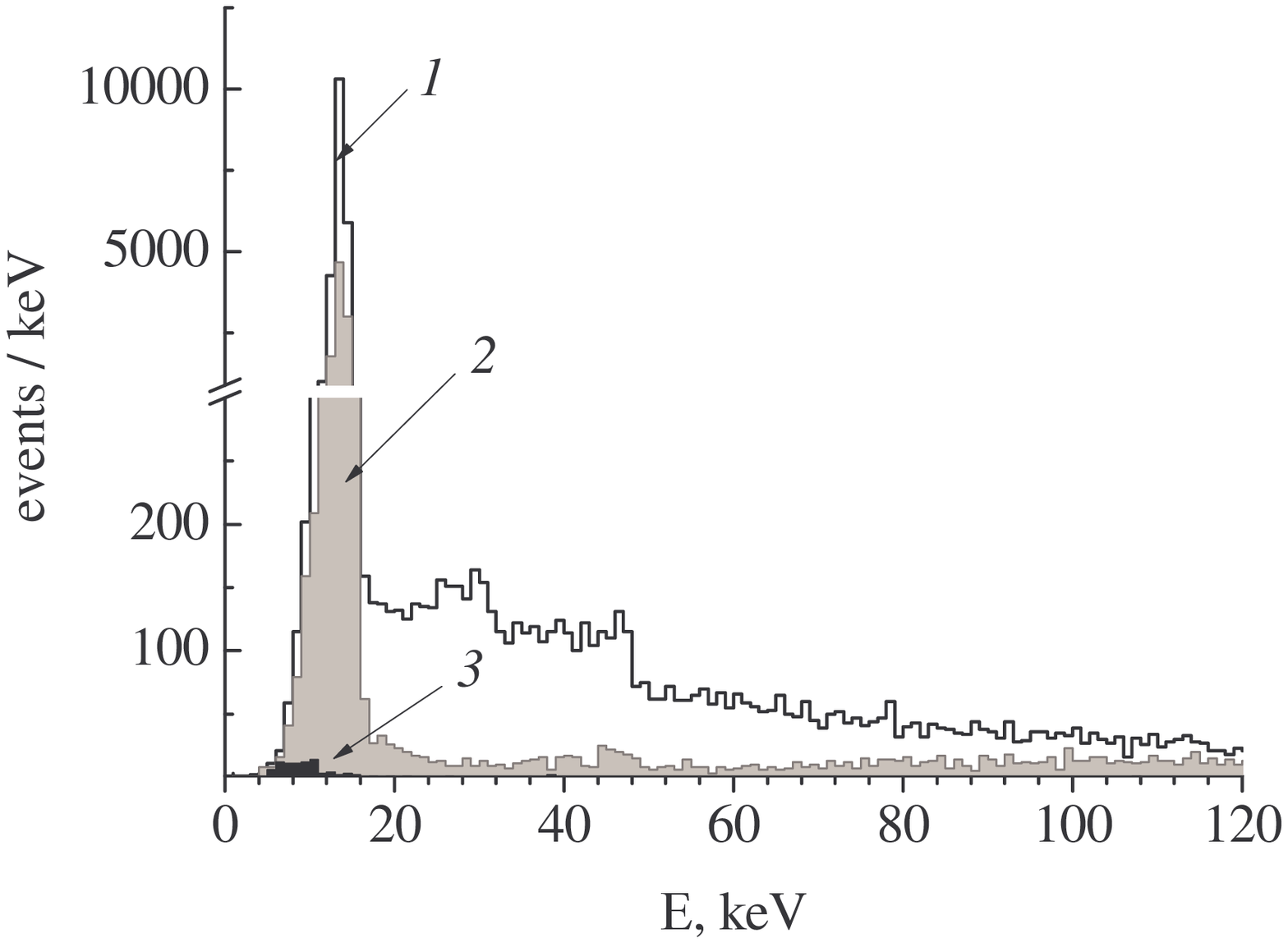}}
\caption{Amplitude spectra (\emph{1}) of single-, (\emph{2})
two-point, and (\emph{3}) three-point events that form the total
analyzed pulse amplitude spectrum of the counter detecting both
background and $^{81}$Kr decays.} \label{spcKr81}
\end{figure}
It decays by electron capture into $^{81}$Br$^*$. In
87.5\% of cases, an electron is captured from the $K$-shell of a Kr
atom ($K$-capture) \cite{r29}. Filling of the vacancy at the
$K$-shell of a daughter Br atom is accompanied in 61.4\% of cases by
an escape of two characteristic quanta with energies of 11.92 keV
($K_{\alpha1}$, 100\%), 11.88 keV ($K_{\alpha2}$, 50.9\%), 13.29 keV
($K_{\beta1}$, 21.0\%), and 13.47 keV ($K_{\beta2}$, 1.07\%)
\cite{r11} and concomitant Auger electrons with energies of 1.55
keV, 1.60 keV, 0.27 keV, and 0.01 keV, respectively (the relative
intensities of $K_{\alpha,\beta}$ lines are presented in brackets).
From this list, it is apparent that, when photons $K_{\alpha1}$ and
$K_{\alpha2}$ escape, the energy of Auger electrons is sufficient
for a distinguishable two-point event to be formed. In 38.6\% of
cases, filling of the vacancy at the $K$-shell of Br is accompanied
by an escape of a cascade of Auger electrons that produce  a
single-point energy deposit. Events with an escape of characteristic
photons $K_{\beta1}$ and $K_{\beta2}$ should also be placed into a
category of single-point events. Taking into account the absorption
efficiency for characteristic photons in the LPC working gas
$(\epsilon_{abs}=0.869)$ and using the above data, one can calculate
the composition of a 13.5-keV full-energy peak: 49.4\% of all events
are single-point, and 50.6\% are two-point.

Describing the noise-free signals by a set of Gaussian curves using
rmse minimization technique (7), one can discriminate between two-
and single-point events. Figure \ref{spcKr81} presents the energy
spectra ({\it 1}) of single-, ({\it 2}) two-~, and ({\it 3})
three-point events forming the total analyzed pulse-amplitude
spectrum of the counter that detects both background and $^{81}$Kr
decays. Similar components (spectra {\it 1-3}) of total spectrum
{\it 1} for a calibration measurement of 88-keV $\gamma$ rays are
shown in Fig.2.

Given the composition of events from the $^{81}$Kr and $^{109}$Cd
sources and results of simulation of these processes in the LPC, one
can, using the GEANT 4.8.2 program \cite{r30}, determine the quality
of suppression of the noise component and the efficiency of the
procedure for separating events by the criteria of a multipoint
event. Table II presents the percentage composition of three types
of events in the full-energy peaks with energies of 13.5 keV
($^{81}$Kr) and 88 keV ($^{109}$Cd), which was obtained by
estimation (column \emph{I}); by simulation of processes in the LPC,
ignoring the confluence of closely spaced primary charge clusters
(column \emph{IIa}), and taking into account their confluence
(column \emph{IIb}); and by separation of the experimental spectrum
into components after its wavelet purification of noise using hard
threshold processing (column \emph{III}).

Determining the parameters of individual components in a compound
event, it is possible to perform energy calibration in the energy
range under investigation and determine the energy resolution of
single-point components which form a multipoint event. The energy
distributions of individual components of two-point events from the
$^{81}$Kr and $^{109}$Cd sources are shown in Figs. 7 and 8,
respectively.
\begin{figure}[!htb]
\resizebox{0.45 \textwidth}{!}{%
\includegraphics{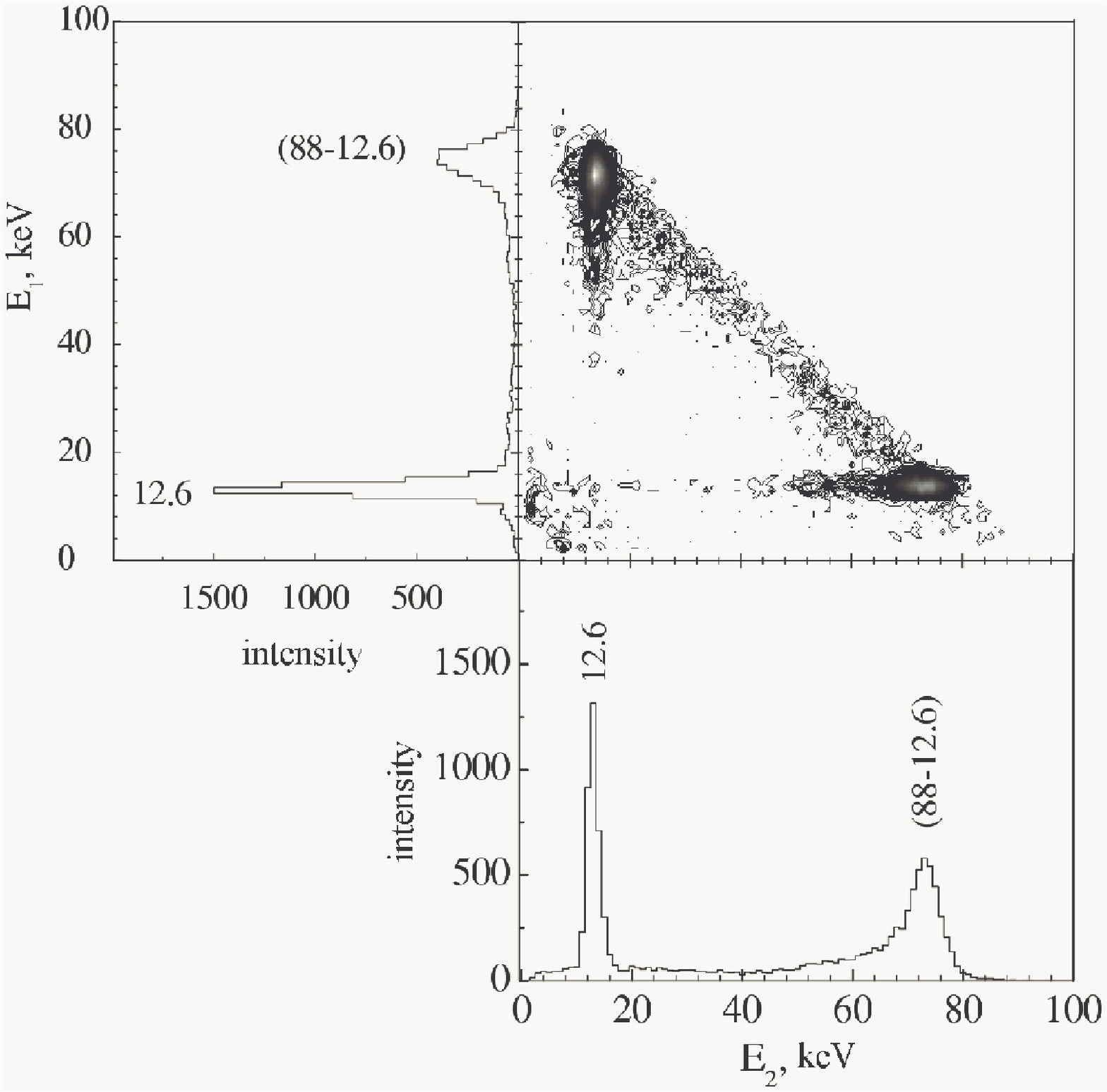}}
\caption{Amplitude distributions of energy deposits  of  individual
components  for two-point events from the source placed in the
middle of the LPC length ($^{109}$Cd).} \label{matrCd109}
\end{figure}

\begin{figure}[!htb]
\resizebox{0.45\textwidth}{!}{%
\includegraphics{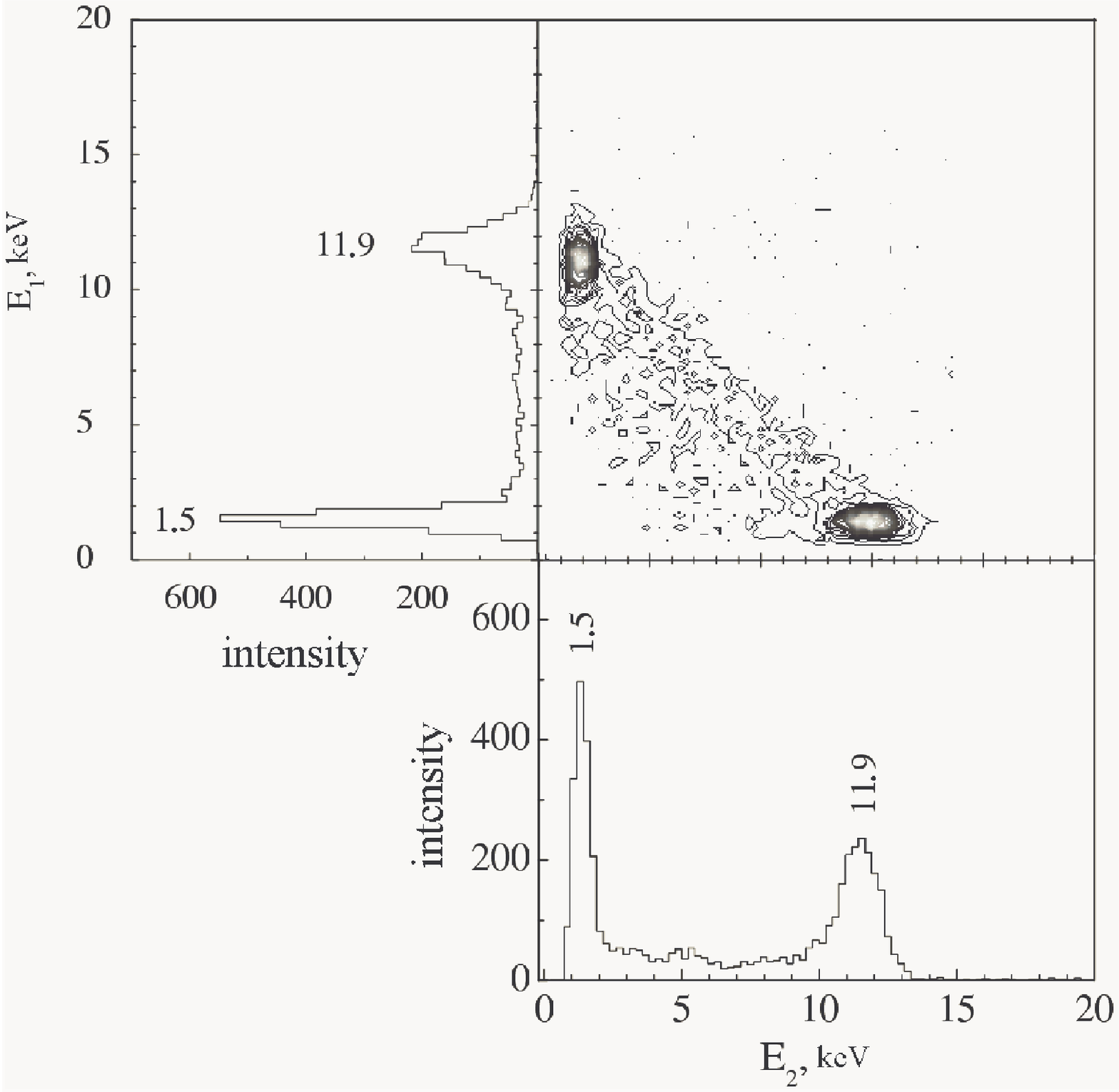}}
\caption{Amplitude distributions of energy deposits in of individual
components of  for two-point events from the internal ($^{81}$Kr)
source with a total energy of 13.5 keV.} \label{matrKr81}
\end{figure}

\section{\label{conclusions}Conclusions} The analysis both of the
pulse shape variety and methods of LPC signal processing has been
performed  using a multiresolutional wavelet analysis. It has been
shown that the noise component can be suppressed successfully, and
the problem of separating of the events with closely spaced
clasters, partially or fully lapping over each other, can be solved
with a high efficiency, and the parameters of these events can be
estimated. The test measurements of signals from the krypton-filled
LPC irradiated by internal and external sources have made it
possible to determine the efficiency of discrimination between
single-, two-, and three-point events. The described technique helps
increase the sensitivity of detection of compound events in
gas-filled detectors and demonstrates its applicability to detection
of rare events from $2K(2\nu)$ capture in large proportional
counters.

\textbf{Acknowledgments}. The work has been carried out under the
financial support of the RFBR (grant no. 04-02-16037) and "Neutrino
Physics" Program of the Presidium of RAS.

\begin{widetext}

\begin{table*}[!htb]
\label{tab2} \caption{Relative composition of three types of events
in the full-energy peaks with energies of 13.5 keV ($^{81}$Kr) and
88 keV ($^{109}$Cd), obtained by estimation (column \emph{I}); by
simulation of processes in the LPC, ignoring the confluence of
closely spaced pointwise ionization regions (column \emph{IIa}) and
taking into account their confluence (column \emph{IIb}); and by
separation of the experimental spectrum into components after its
wavelet purification of noise using hard threshold processing
(column \emph{III}).}
\begin{tabular}{|c|c|c|c|c||c|c|c|c|}
  \hline
& \multicolumn{4} {c||}{13.5 keV ($^{81}$Kr)} & \multicolumn{4}{c|}
{88 keV ($^{109}$Cd)} \\ \hline &    calc., \% & \multicolumn{2}
{c|} {GEANT4.8.2, \%} &  \multicolumn{1} {c||} {empiric., \%}   &
calc., \% & \multicolumn{2} {c|} {GEANT4.8.2, \%}   &
\multicolumn{1} {c|} {empiric., \%}
           \\ %\hline
type of events &\emph{I}&\emph{IIa}&\emph{IIb}&\emph{III}
&\emph{I}&\emph{IIa}&\emph{IIb}&\emph{III} \\ \hline\hline
  singl-point &  49.4 & 44.1 & 66.2 & 68.2 & 44.1 & 45.0 & 59.4 & 56.8  \\
  two-point   &  50.6 & 55.9 & 33.8 & 31.7 & 53.4 & 53.7 & 40.0 & 39.4  \\
  three-point &       &      &      & $<0.1$ &  2.5 &  1.3 &  0.6 &  3.4  \\
  \hline
\end{tabular}

\end{table*}

\end{widetext}

%%%%%%%%%%%%%%%%%%%%%%%%%%%%%%%%%%%%%%%%%%%%%%%%%%%%%%%%%%%%%%%%%%%%%%%%%%%%%%%%%%%%%%%

% reff ----------------------------------------------------------------------------------------

%%%%%%%%%%%%%%%%%%%%%%%%%%%%%%%%%%%%%%%%%%%%%%%%%%%%%%%%%%%%%%%%%%%%%%%%%%%%%%%%%%%%%%

\end{document}